\pdfoutput=1
\documentclass[aps,pre,10pt,twocolumn,floats,floatfix,showpacs,showkeys,superscriptaddress]{revtex4}
\usepackage{graphicx,amssymb,amsmath}
\usepackage[latin1]{inputenc}
\usepackage{color}

\definecolor{red}{rgb}{1,0,0}
\definecolor{green}{rgb}{0,1,0}
\definecolor{blue}{rgb}{0,0,1}
\definecolor{darkmagenta}{rgb}{.5,0,.5}

\begin{document}
\title[Oscillator]{Stochastic dynamics and the noisy Brusselator behaviour}

\author{Nicol{\'a}s Rubido}
\affiliation{Institute for Complex Systems and Mathematical Biology, University of Aberdeen,
	     King's College, AB24 3UE Aberdeen, United Kingdom}
\affiliation{Instituto de F{\'i}sica, Facultad de Ciencias, Universidad de la Rep{\'u}blica,
	     Igu{\'a} 4225, 11200 Montevideo, Uruguay}

\date{\today}
\begin{abstract}
In non-linear dynamics there are several model systems to study oscillations. One iconic
example is the ``Brusselator'', which describes the dynamics of the concentration of two
chemical species in the non-equilibrium phase. In this work we study the Brusselator
dynamics as a stochastic chemical reaction without diffusion analysing the corresponding
stochastic differential equations with thermal or multiplicative noise. In both stochastic
scenarios we investigate numerically how the Hopf bifurcation of the non-stochastic system
is modified. Furthermore, we derive analytical expressions for the noise average orbits and
variance of general stochastic dynamics, a general diffusion relationship in the thermal
noise framework, and an asymptotic expression for the noise average quadratic deviations.
Hence, besides the impact of these results on the noisy Brusselator's dynamics, our findings
are also relevant for general stochastic systems.
\end{abstract}

\maketitle
\section{Introduction}
There are many instances in which physical systems spontaneously become emergent or orderly
\cite{Strogatz,Murray,Prigogine}. Even more spectacular is the order
created by chemical systems; the most dramatic being the order associated with life. However,
not all chemical reactions generate order. The class of reactions most closely associated
with order creation are the auto-catalytic reactions. These are chemical reactions in which
at least one of the reactants is also a product, hence, the equations are fundamentally
non-linear due to this feedback effect.

Simple auto-catalytic reactions are known to exhibit sustained oscillations \cite{Prigogine,
Broomhead,Osipov}, thus, creating temporal order. Other reactions can generate separation
of chemical species generating spatial order, i.a., the Belousov-Zhabotinsky reaction
\cite{Agladze}. More complex reactions are involved in metabolic pathways and networks in
biological systems \cite{Petsko,Etay,Hilborn,Ciandrini}. The transition to order as the
distance from equilibrium increases is not usually continuous. Order typically appears
abruptly. The threshold between the disorder of chemical equilibrium and order happens as
a phase transition. The conditions for a phase transition to occur are determined with the
mathematical machinery of non-equilibrium statistical mechanics \cite{Reichl}.

A paradigmatic example of an auto-catalytic reaction, which exhibits out of equilibrium
oscillations, is the ``\emph{Brusselator}''. It describes the dynamics of the concentration
of two chemical species, where the evolution of each component is obtained from the
following dimensionless differential equations \cite{Prigogine,Murray,Broomhead,Osipov}
\begin{equation}
 \left\lbrace \begin{array}{lcl}
               d\,u/d\,\tau & = & 1 - \left( b + 1 \right)\,u + a\,u^2\,v = f\left(u,\,v
\right)\,,\\
	       d\,v/d\,\tau & = & b\,u - a\,u^2\,v = g\left(u,\,v\right)\,,
              \end{array} \right.
 \label{eq_oscillator}
\end{equation}
where $a,\,b > 0$ are constants, $u$ and $v$ correspond to the concentrations of the two
species, and $\tau$ is the dimensionless time. These differential equations show, for most
initial conditions, self-sustained oscillations when $b > 1 + a$. The transition from the
chemical equilibrium state to this oscillatory behaviour happens via a Hopf bifurcation
\cite{Guckenheimer}.

Classical and recent work has emphasized the importance of fluctuations to better model the
internal evolution of macroscopic systems which are only accounted for by stochastic models
\cite{Broomhead,Osipov,Agladze,Petsko,Traulsen,Jan,Melbinger,Beta,Abbott}. In particular,
Chemical oscillations might be augmented and/or disturbed by stochastic effects and random
drift. For instance, models of diffusion-driven pattern formation that rely on the Turing
mechanism are commonly used in science. Nevertheless, many such models suffer from the
defect of requiring fine tuning of parameters in order to predict the formation of spatial
patterns. The limited range of parameters for which patterns are seen could be attributed
to the simplicity of the models chosen to describe the process; however, for systems with
an underlying molecular basis another explanation has recently been put forward
\cite{Biancalani,Biancalani2,Butler}. Those authors have observed that Turing-like patterns
exist for a much greater range of parameter values if the discrete nature of the molecules
comprising the system is taken into account. The systems within this class may be analysed
using the theory of stochastic processes. For the Brusselator, the inclusion of noise
affects the concentrations of the relevant chemical species and accounts for the molecular
character of the reaction compounds.

In this work, we study the Brusselator system without diffusion as a stochastic process due
to the inclusion of thermal or multiplicative noise. We discuss the non-stochastic dynamics
of the Brusselator analytically (Sec.~\ref{sec_brusselator}) and derive a general expression
for the noise average linear and quadratic deviations of the thermal noise stochastic system
(Sec.~\ref{sec_stochastic}) from generic Stochastic Differential Equations (SDEs).
Consequently, we obtain a general Einstein diffusion relationship (Sec.~\ref{sec_diffusion})
and derive a general expression for the asymptotic behaviour of the quadratic deviations.
These expressions are derived in terms of the eigenvalues and eigenvectors of the Jacobian
matrix of the non-stochastic equations. Furthermore, we analyse how the Brusselator Hopf
bifurcation is perturbed by thermal or multiplicative noise via numerical experiments
(Sec.~\ref{sec_Hopf}). Our results show that, in both frameworks, the bifurcation transition
is kept in average for small noise intensities ($\Gamma \lesssim 0.1$). Our findings are
relevant, not only for the analysis of the noisy Brusselator chemical reaction, but also
for general stochastic systems.

\section{Model}

 \subsection{Dynamical features of the Brusselator}
  \label{sec_brusselator}
The equilibrium phase of the Brusselator chemical reaction is given by the fixed point (FP)
of Eq.~(\ref{eq_oscillator}), namely, $\left(u_0,\,v_0\right) = \left(1,\,b/a\right)$. The
Jacobian of Eq.~(\ref{eq_oscillator}) evaluated at the FP is given by
\begin{equation}
  D\vec{F}_{(u_0,\,v_0)} = \left( \begin{array}{cc}
                    b - 1 &  a \\
                    -b   & -a
                   \end{array} \right),
 \label{eq_jacobian}
\end{equation}
where each entry of the matrix corresponds to the partial derivatives given by $\left(
\partial\,f_i/\partial\,x_j\right)_{(u_0,\,v_0)}$, $f_i$ being $f_1 = f$ or $f_2 = g$,
with $x_1 = u$ and $x_2 = v$. The eigenvalues $\lambda_\pm$ of Eq.~(\ref{eq_jacobian})
are the solutions of the characteristic polynomial $\chi\left(\lambda\right) = \det\left(
D\vec{F}_{u_0,\,v_0} - \lambda\,\mathbf{I}\right) = 0$, i.e.,
\begin{equation}
  \lambda_{\pm} = \frac{\mathsf{Tr}}{2} \pm \sqrt{ \frac{\mathsf{Tr}^2}{4} - \Delta }\,,
 \label{eq_eigenvalues}
\end{equation}
and determine the local stability of the FP. As $\Delta = a > 0$, $\Delta$ being the
determinant of Eq.~(\ref{eq_jacobian}), a saddle node FP is impossible \cite{Guckenheimer}.
Hence, analysing the sign of the trace [$\mathsf{Tr} = b - \left(1 + a\right)$] of
Eq.~(\ref{eq_jacobian}), we have that the FP is \emph{unstable} if $\mathsf{Tr} > 0$ and
that the FP is \emph{stable} if $\mathsf{Tr} < 0$. The \emph{critical point} is then $b_c
\equiv 1 + a$.

\begin{figure}[htbp]
 \begin{center}
  \includegraphics[width=0.95\columnwidth]{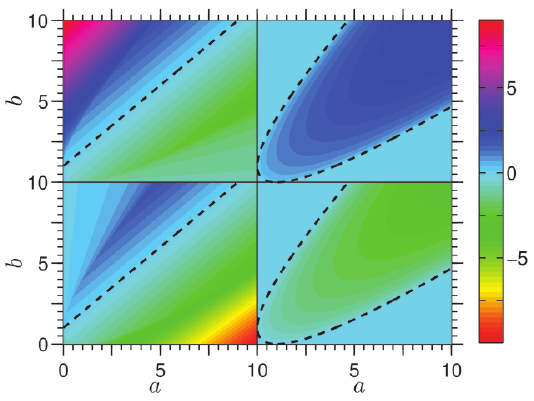}
 \end{center} \vspace{-1pc}
  \caption{(Color online) Real (left column) and imaginary (right column) parts of the
  eigenvalue solutions ($\lambda_\pm$ in colour code) for the Brusselator Jacobian matrix at
  the fixed point (FP). The top row panels show $\lambda_{+}$ and the bottom row panels
  $\lambda_{-}$ [positive and negative solutions of Eq.~(\ref{eq_eigenvalues}),
  respectively]. The dashed lines on the left column panels indicate the critical border
  $b_c = 1 + a$. The FP is stable for values beneath the dashed line, otherwise, is unstable.
  The dashed lines on the right column panels indicate the border between ordinary (outside
  the dashed area) and spiral (inside the dashed area) FP.}
 \label{fig_eigenvalues}
\end{figure}

The square root in Eq.~(\ref{eq_eigenvalues}) determines if the FP is ordinary ($\lambda_{
\pm}\in\mathbb{R}$) or spiral ($\lambda_{\pm}\in\mathbb{C}$). Thus, two two-fold cases
appear, as they are shown in Fig.~\ref{fig_eigenvalues} (colour code). From the left column
panels in Fig.~\ref{fig_eigenvalues}, it is seen that the line $b_c = 1 + a$ (diagonal
dashed line) divides the parameter space in an upper positive region ($\mathsf{Re}\{
\lambda_{\pm}\} > 0$) and a lower negative region ($\mathsf{Re}\{\lambda_{\pm}\} < 0$).
These two regions account for the unstable and stable FP situations, respectively. On the
right column of the figure, the spiral and ordinary characteristics of the FP eigenvalues
are discriminated by the critical set $\mathsf{Tr}^2 = 4\Delta$ (curved dashed line). The
outer region of this set has the imaginary part null, namely, $\mathsf{Im}\{\lambda_{\pm}
\} = 0$, thus, the FP is ordinary classified. On the inside region, the FP is spiral
(unstable, if it is above the critical level, and stable if it is below).

\begin{figure}[htbp]
 \begin{center}
  \includegraphics[width=1.0\columnwidth]{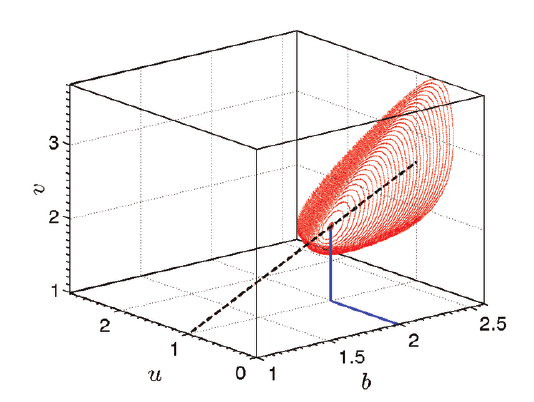}
 \end{center} \vspace{-1pc}
  \caption{(Color online) Brusselator's bifurcation diagram as $b$ is increased for $a = 1$.
  Red dots correspond to the system's asymptotic orbits. The critical point $b_c = 1 + a =
  2$ is signalled by straight (blue online) lines. The dashed (black online) line corresponds
  to the steady state solution $\left(1,\,b/a\right)$, which becomes unstable after the
  critical point as a Hopf bifurcation.}
 \label{fig_bifurcation}
\end{figure}

However, the main feature for self-sustained oscillations to occur is to have an unstable
FP ($b > 1 + a$). In that case, solutions are attracted to a limit-cycle \cite{Guckenheimer,
Jan}. A limit-cycle on a plane (or a two-dimensional manifold) is a closed trajectory in
phase space having the property that at least one other trajectory spirals into it either
as time approaches infinity or as time approaches negative infinity \cite{Guckenheimer}.
This behaviour is shown in Fig.~\ref{fig_bifurcation} for the Brusselator. For $b < b_c$ the
FP is stable and there is no self-sustained oscillation. Such a steady state solution
becomes unstable under a Hopf bifurcation as $b$ is increased. A stable oscillation appears
for $b > b_c$, which corresponds to the non-equilibrium phase of the system.

 \subsection{Stochastic Dynamics}
A general coupled set of first order Stochastic Differential Equations (SDEs) is given by
\begin{equation}
  \dot{\vec{x}}_\eta(t) = \vec{F}\left(\vec{x}_\eta(t),\,t\right) +
   \mathbf{H}\left(\vec{x}_\eta(t),\,t\right)\,
   \vec{\eta}(t)\,,
 \label{eq_SDEs}
\end{equation}
where we name $\vec{x}_\eta = \{ x_i:\;i = 1,\,\ldots,\,N\}$ as the set of state variables,
$\vec{F}$ as the deterministic part of the SDE (namely, a field vector $\vec{F}$ with
components $f_i:\mathbb{R}^N\times \mathbb{R} \to \mathbb{R}$ known as the \emph{drift
coefficients}), $\mathbf{H}$ as the $N \times N$ coupling matrix with function entries
$h_{ij}:\mathbb{R}^N\times\mathbb{R} \to \mathbb{R}$ [i.e., $h_{ij}\left(\vec{x}_\eta(t),\,
t\right)$], known as the \emph{diffusion coefficients}, and $\vec{\eta}$ as the vector
of random fluctuations. The noise is assumed to be uncorrelated and with zero mean for
all coordinates, namely,
\begin{eqnarray}
\nonumber
 \left\langle \eta_i(t)\,\eta_j(s) \right\rangle = \delta_{ij}\,\delta\left(t - s\right)\,,
 \\ \text{and}\;\;
 \left\langle \eta_i(t) \right\rangle = 0\,,\;\forall\,i = 1,\,\ldots,\,N\,,
 \label{eq_noise}
\end{eqnarray}
where, as in the following, we denote $\left\langle \cdots \right\rangle$ to be the average
over various noise realisations (in other words, a mean value over an ensemble of possibly
different fluctuations), $\delta_{ij}$ to be the Kronecker delta, and $\delta(t-s)$ to be
the Dirac delta function.

The system is said to be subject to \emph{additive or thermal noise} if the drift
coefficients ($h_{ij}$) are constants, otherwise it is said to be subject to
\emph{multiplicative noise}. For the Brusselator, taking into account random fluctuations
in the equations of motion modifies the concentrations of the relevant chemical species into
stochastic variables. The inclusion of noise is intended to account for the molecular
character of the chemical reaction. Hence, the \emph{Stochastic Brusselator Equations}
(SBEs) are
\begin{equation}
  \left[\! \begin{array}{c}
          \dot{u} \\ \dot{v}
         \end{array} \!\right] = \left[\! \begin{array}{c}
				       f\left(u,\,v\right) \\ g\left(u,\,v\right)
				      \end{array} \!\right] + \mathbf{H}\left(u,\,v\right)
				  \left[\! \begin{array}{c}
				\eta_{u} \\ \eta_{v}
					\end{array} \!\right]\,,
 \label{eq_BLEs}
\end{equation}
with
\begin{equation}
  \mathbf{H}\left(u,\,v\right) = \left[\! \begin{array}{cc}
				h_{11}\!\left(u,\,v\right) & h_{12}\!\left(u,\,v\right) \\
				h_{21}\!\left(u,\,v\right) & h_{22}\!\left(u,\,v\right)
				  \end{array} \!\right]\,,
 \label{eq_coupling}
\end{equation}
where the deterministic drift coefficients come from Eq.~(\ref{eq_oscillator}) and the
chemical concentrations $u$ and $v$ are now stochastic variables.

In particular, the analysis of how the stability of the FP (sub-critical parameter values)
and the limit-cycle (above criticality) of the deterministic Brusselator changes due to the
inclusion of noise is carried by linearising the SBEs around the particular stable solution.
We tackle this analysis on the general SDE case [Eq.~(\ref{eq_SDEs})]. We start by analysing
the effect of the noise over the stability of the FP and \emph{find that the resulting
average solutions are mainly affected by the eigenvalues and eigenvectors of the Jacobian
of the field vector $\vec{F}$ and the drift coefficients}.

Let $\delta\vec{x}_{\eta} \equiv\vec{x}_{\eta} - \vec{x}_{eq}$ be the \emph{deviation}
vector, where $\vec{x}_{eq}$ is the steady equilibrium solution (FP). Assuming that this
deviation is small for all times and any noise realisation, then,
\begin{eqnarray}
 \nonumber
  \delta\dot{\vec{x}}_{\eta} = \left[D\vec{F}_{\vec{x}_{eq}} + \left( \nabla\left[
   \mathbf{H}\left(\vec{x}_\eta(t),\,t\right) \,\vec{\eta}(t)\right] \right)_{\vec{x}_{eq}}
    \right]\times \\ \times\delta\vec{x}_{\eta} +
  \mathbf{H}_{\vec{x}_{eq}}\,\vec{\eta}(t)\,,
 \label{eq_linear-BLEs}
\end{eqnarray}
$D\vec{F}_{\vec{x}_{eq}}$ ($\mathbf{H}_{\vec{x}_{eq}}$) being the Jacobian (coupling) matrix
evaluated at the FP and $\nabla$ the gradient operator. Unless the noise is additive, the
matrices inside the square brackets have a noise dependence. Thus, we restrict ourselves to
the case of thermal noise for the mathematical derivations.

For \emph{additive noise}, $\mathbf{H}(\vec{x}_\eta(t),t)$ is a constant matrix,
$\mathbf{H}$, independent of the system's state vector and time, hence, direct integration
of Eq.~(\ref{eq_linear-BLEs}) is possible. Denoting $D\vec{F}_{\vec{x}_{eq}} = \mathbf{J}$,
the deviation vector evolves according to
\begin{equation}
  \delta\vec{x}_{\eta}(t) = e^{\mathbf{J}\,t}\delta\vec{x}(0) + \int_0^t ds\;e^{\mathbf{J}
   \,(t - s)}\,\mathbf{H}\,\vec{\eta}(s)\,.
 \label{eq_BLEs-linear_sol}
\end{equation}

We note that the exponentials of the Jacobian matrix in the former expressions are
understood to be a matrix exponent, thus, they are computed in a power series expansion
using its spectral decomposition [$\mathbf{J} = \mathbf{P}\mathbf{\Lambda}\mathbf{P}^{-1}$,
where $\left(\mathbf{P}\right)_{ij} = \left(\vec{v}_j\right)_i$ is the $i$-th coordinate of
the $j$-th eigenvector of $\mathbf{J}$ and $\left(\mathbf{\Lambda}\right)_{ij} = \delta_{ij}
\lambda_j$, with $\lambda_j$ being the $j$-th eigenvalue of $\mathbf{J}$]. Specifically,
\begin{equation}
  e^{\mathbf{J}\,t} \equiv \sum_{n = 0}^\infty \frac{1}{n!}\left(\mathbf{J}\,t\right)^n =
   \mathbf{P}\,e^{\mathbf{\Lambda}\,t}\,\mathbf{P}^{-1}\,.
 \label{eq_spectral}
\end{equation}

\section{Results}

 \subsection{Thermal noise ensemble first and second moments}
  \label{sec_stochastic}
Due to the stochastic character of Eq.~(\ref{eq_BLEs-linear_sol}), we focus on the
analytical derivation of the first and second moments of the thermal noise ensemble
distribution. In other words, we derive an expression for the noise average value of the
deviations with respect to the FP and the noise average value of the quadratic deviations
with respect to the FP. The noise average value of the deviations, $\vec{m}(t) \equiv
\left\langle \delta\vec{x}_{\eta}(t) \right\rangle$, is
\begin{equation}
  \vec{m}(t) = e^{\mathbf{J}\,t}\,\delta\vec{x}(0) + \int_0^t ds\;e^{\mathbf{J}\,( t - s )}
   \,\left\langle \mathbf{H}\,\vec{\eta}(s) \right\rangle\,,
 \label{eq_mean_BLE-sol}
\end{equation}
and the noise average value of the quadratic deviations, $\rho^2(t) \equiv \left\langle
\left[\delta\vec{x}_{\eta}(t)\right]^2 \right\rangle$, is
\begin{eqnarray}
 \nonumber
  \rho^2(t) =  e^{\mathbf{J}\,t}\,\delta\vec{x}(0) \cdot e^{\mathbf{J}\,t}\,\delta\vec{x}(0)
  + \\
 \nonumber
    2e^{\mathbf{J}\,t}\,\delta\vec{x}(0) \cdot \int_0^t ds\;e^{\mathbf{J}\,( t - s )}
   \,\left\langle \mathbf{H}\,\vec{\eta}(s) \right\rangle + \\
  \int_0^t\!ds\!\int_0^s\!ds' \left\langle e^{\mathbf{J}( t - s)}\,\mathbf{H}\,\vec{\eta}(s)
   \cdot e^{\mathbf{J}( t - s' )}\,\mathbf{H}\,\vec{\eta}(s') \right\rangle,
 \label{eq_std_BLE-sol}
\end{eqnarray}
where `` $\cdot$ '' is the inner product between vectors. In particular, for the Jacobian
eigenvectors we have that $\vec{v}_n \cdot \vec{v}_m = \sum_{i=1}^N \left(\vec{v}_n\right)_i
\left(\vec{v}_m\right)_i^{\star} = \delta_{nm}$, `` $^\star$ '' being the complex conjugate
operation.

As the random fluctuations are additive, then $\mathbf{H}$ is independent of the noise
realisation, hence, $\left\langle\mathbf{H}\eta\right\rangle = \mathbf{H} \left\langle
\eta\right\rangle = 0$. Consequently, the \emph{noise average value of the deviations}
evolves as
\begin{equation}
  \vec{m}(t) = \mathbf{P}\,e^{\mathbf{\Lambda}\,t}\,\mathbf{P}^{-1}\,\delta\vec{x}(0)\,,
 \label{eq_mean}
\end{equation}
which is the first main result of this work and constitutes the first moment of the thermal
noise ensemble orbits. Equation~(\ref{eq_mean}) says that the average of the deviation
variables tends to zero for $t \to \infty$ if and only if the eigenvalues correspond to a
stable FP, namely, when $\mathsf{Re}\{{\lambda_i}\} < 0 \;\forall\,i$. Hence, \emph{the
stochastic system returns to the steady state (in average) after being perturbed}, as long
as the noise intensities are small (i.e., $\left\|\delta\vec{x}_{\eta}(t)\right\| \ll 1$
$\forall\,t$).

\begin{figure}[htbp]
 \begin{center}
  \begin{minipage}{10pc}
   \textbf{(a)}\\
   \includegraphics[width=10pc]{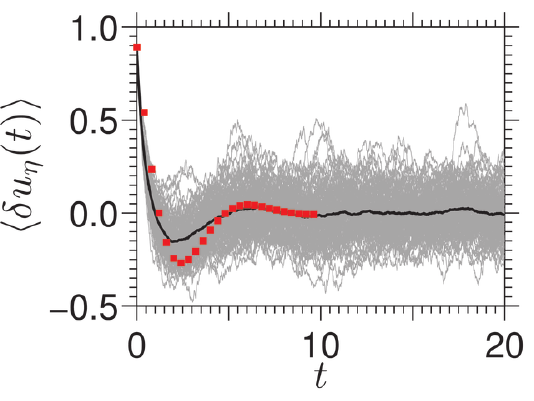}
  \end{minipage}
  \begin{minipage}{10pc}
   \textbf{(b)}\\
   \includegraphics[width=10pc]{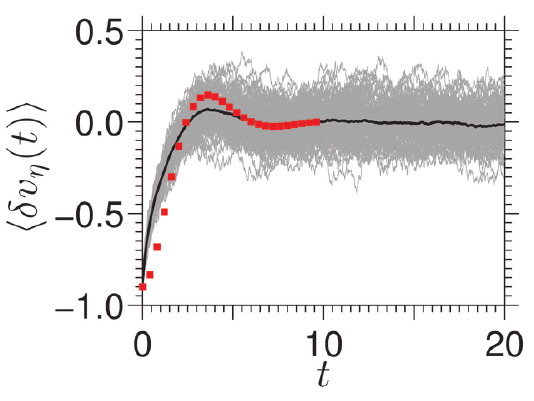}
  \end{minipage}
 \end{center} \vspace{-1pc}
  \caption{(Color online) Panel {\bf (a)} [Panel {\bf (b)}] shows $100$ stochastic orbits
  of the Brusselator's chemical concentration deviations from the fixed point $\delta u_\eta$
  [$\delta v_\eta$] in continuous light (grey online) curves. The diffusion coefficients are
  constant (thermal noise scenario) and given by $h_{ij} = \delta_{ij} \Gamma$, with $\Gamma
  = 10^{-1}$. The noise average orbit deviations in these panels are represented by dark
  (black online) continuous curves. The analytical predictions [Eq.~(\ref{eq_mean})] for
  these averages are shown by filled (red online) squares. Both panels corresponds to the
  sub-critical regime, with $a = b = 1$.}
 \label{fig_BLE_orbits}
\end{figure}

In particular, for the Brusselator, the analytical result of Eq.~(\ref{eq_mean}) predicts
how the noisy chemical concentrations converge to the FP if averaged over various thermal
noise realisations, as it is demonstrated in Fig.~\ref{fig_BLE_orbits}. Both panels show
how the analytical prediction (filled squares) of Eq.~(\ref{eq_mean}) for $\vec{m}(t) =
\left( \left\langle \delta u_\eta(t) \right\rangle,\,\left\langle \delta v_\eta(t)
\right\rangle \right)$ has a remarkable agreement with the numerical experiments, although
the initial perturbation is large [the initial condition for every stochastic orbit is
$\left(\delta u_\eta(0),\, \delta v_\eta(0) \right) = \left(0.9,\,-0.9\right)$].

On the other hand, the \emph{noise average value of the quadratic deviations} of the SDE
orbits around the steady equilibrium state evolve as
\begin{eqnarray}
 \nonumber
  \rho^2(t) = \delta\vec{x}(0)\cdot \mathbf{P}\,e^{2\mathsf{Re}\{\mathbf{\Lambda}\}\,t}\,
   \mathbf{P}^{-1} \delta\vec{x}(0) + \\
  \sum_{i,j,k,n}^N h_{ik}\left(\vec{v}_n\right)_i \left[ \frac{e^{2\,\mathsf{Re}\{\lambda_n
   \}\,t} - 1}{ 2\,\mathsf{Re}\{\lambda_n\} }\right]\left(\vec{v}_n\right)_j^\star h_{jk}\,,
 \label{eq_std}
\end{eqnarray}
where $\left(\vec{v}_n\right)_{i}$ is the $i$-th coordinate of the eigenvector associated to
the $\lambda_n$ eigenvalue of the Jacobian matrix in the particular steady state solution
and all summations run from $1$ to $N$.

Equation~(\ref{eq_std}) is the second main results in this work and constitutes the second
moment of the ensemble of thermal noise realisations. It is derived by applying direct
integration of Eq.~(\ref{eq_std_BLE-sol}), the noise properties defined in
Eq.~(\ref{eq_noise}), and the spectral decomposition of the Jacobian matrix
[Eq.~(\ref{eq_spectral})]. It expresses how the noise and the deterministic part of the SDE
produce divergence (or convergence) in the trajectories of neighbouring initial conditions
over the ensemble of thermal noise realisations close to the FP solution. The first right
hand side term in Eq.~(\ref{eq_std}) diverges (converges) if the FP is unstable (stable).
In such case, neighbouring initial conditions are driven away (closer) at a rate given by
the real part of the exponents of the system, i.e., by $2\mathsf{Re}\{\lambda_n\}$. The
second term on the right hand side of Eq.~(\ref{eq_std}) accounts for the divergence
(convergence) due to the stochasticity in the system.

In order to find the \emph{variance} $\sigma^2(t)$ of the SDE, it is enough to discard the
first term on the right side of Eq.~(\ref{eq_std}). Specifically,
$$
  \sigma^2(t) = \rho^2(t) - \vec{m}(t)\cdot\vec{m}(t) = \left\langle \left[\vec{x}_\eta(t)
   \right]^2 \right\rangle - \left[ \left\langle \vec{x}_\eta(t) \right\rangle \right]^2\,,
$$
\begin{equation}
  \sigma^2(t) = \sum_{i,j,k,n}^N{\left(\vec{v}_n\right)_i h_{ik}\! \left[ \frac{e^{2\,
   \mathsf{Re}\{\lambda_n\}\,t} - 1}{ 2\,\mathsf{Re}\{\lambda_n\} }\right]\!
    \left(\vec{v}_n\right)_j^\star h_{jk} }.
 \label{eq_variance}
\end{equation}

  \subsection{Thermal noise diffusion relationship and the variance asymptotic behaviour}
   \label{sec_diffusion}
Our general diffusion relationship is derived from Eq.~(\ref{eq_variance}) by considering
the small time-scales. In such transient window, an expansion in power series up to the
first order results in the following \emph{Einstein diffusion relationship}, $\sigma^2(t)
\simeq D\,t$,
\begin{equation}
  D \equiv \sum_{i,j}^N h_{ij}^2 = \mu\,k_B T\,,
 \label{eq_Einstein_rel}
\end{equation}
where $\mu$ is the mobility coefficient, $k_B$ is Boltzmann's constant, and $T$ is the
temperature.

It is worth mentioning that Eq.~(\ref{eq_Einstein_rel}) corresponds to the rate at which the
variance of the perturbations grows in time averaged over the various random fluctuation
realisations. Hence, it may not be directly relatable to the regular Brownian Motion (BM)
solution under an external potential in the over-damped regime, i.e., the known Einstein
diffusion relationship \cite{Reichl}.

On the one hand, the diffusion relationship that BM achieves corresponds to the particle's
position variance, though it depends on the fact that the external force is derived from a
potential. However, if the Brusselator field vector ($\vec{F}$) is derived from a
bi-dimensional potential, periodic solutions are absent ($\nabla\times\vec{F} = 0$),
such as the limit-cycle state. On the other hand, the variance relationship for the BM,
that holds the Einstein's diffusion relationship, is a relationship regarding how
much the particle diffuses as if it performed random walks as a function of the temperature,
i.e., the parameter that regulates the ``strength'' of the fluctuations. For the Brusselator,
the relationship given by Eq.~(\ref{eq_Einstein_rel}), predicts a similar behaviour for the
variance of the perturbations in the system out of the equilibrium state when subject to
additive noise, but there is no particle movement involved. Instead, the ``movement''
corresponds to the chemical concentration fluctuations. Consequently, the general diffusion
relationship we find is only mentioned as a qualitative Einstein diffusion relationship
analogue which exhibits the same mathematical formulation as the one for BM.

For the Brusselator, in the case where the thermal noise only affects each chemical
concentration independently, namely, $h_{ij} = \delta_{ij}\Gamma$, then $\sum_{i,j}^2
h_{ij}^2 = 2\Gamma^2$, and Eq.~(\ref{eq_Einstein_rel}) results in
\begin{equation}
  D = 2\Gamma^2 = \mu\,k_B T\,,
 \label{eq_Einstein_rel_ex}
\end{equation}
where we can continue the analogy with the BM and say that the $2$ appearing in the
relationship corresponds to the $2$ degrees of freedom of the Brusselator.

For a stable FP, the variance [Eq.~(\ref{eq_variance})] saturates at a finite value,
meaning that initially neighbouring orbits are found always within a fixed distance from
the FP. In other words, for a stable FP, the variance of the thermal noise SDE converges
to
\begin{eqnarray}
  \lim_{t\to\infty} \sigma^2(t) = 
  -\sum_{i,j,k,n}^N \left(\vec{v}_n\right)_i h_{ik} \frac{1}{ 2\,\mathsf{Re}\{\lambda_n\} }
    \left(\vec{v}_n\right)_j^\star h_{jk}\,,
 \label{eq_std_asympt}
\end{eqnarray}
with $\mathsf{Re}\{\lambda_n\} < 0\;\forall\,n$. This constitutes the final analytical
result of this work.

\begin{figure}[htbp]
 \begin{center}
  \begin{minipage}{10pc}
   \textbf{(a)}\\
   \includegraphics[width=10pc]{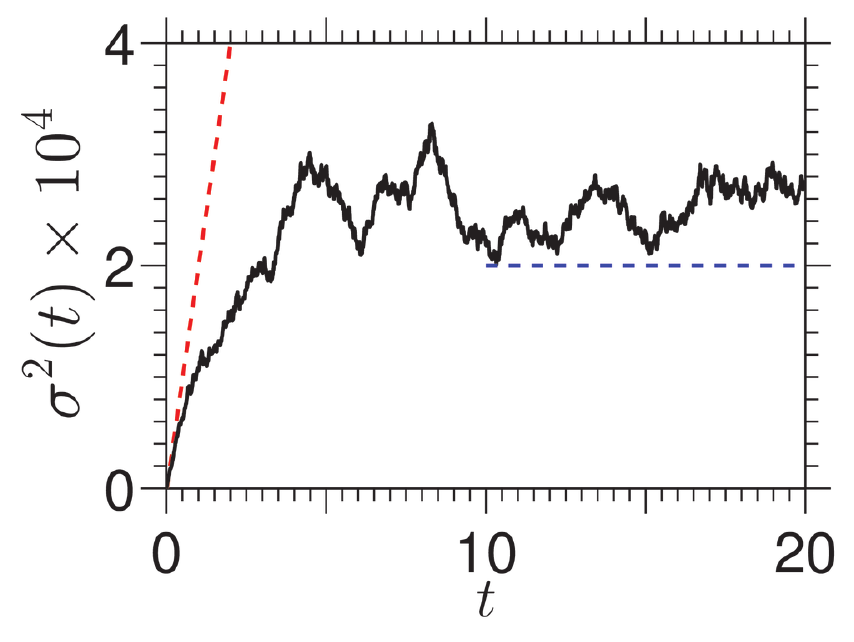}
  \end{minipage}
  \begin{minipage}{10pc}
   \textbf{(b)}\\
   \includegraphics[width=10pc]{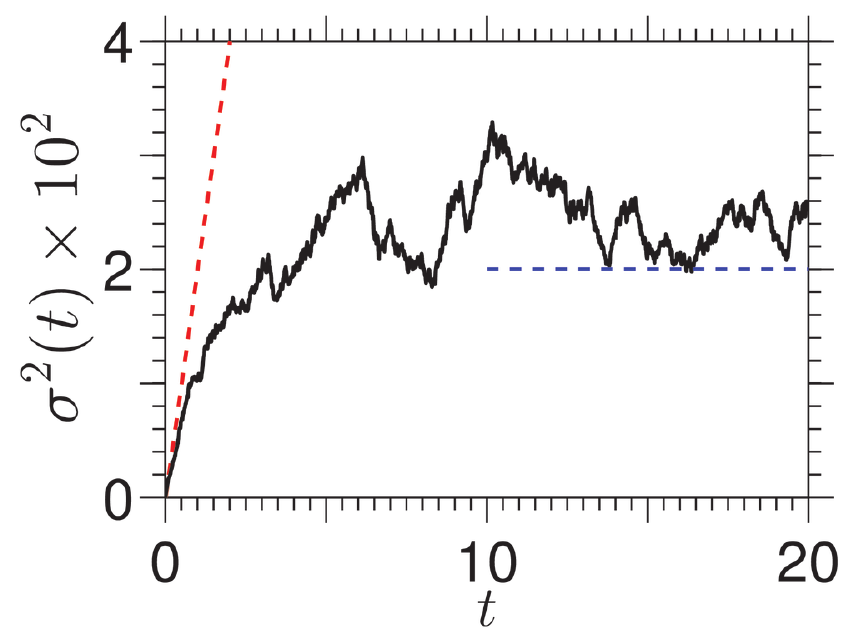}
  \end{minipage}
 \end{center} \vspace{-1pc}
  \caption{(Color online) The panels show the sub-critical ($a = b = 1$) noise average
  orbits variance, namely, $\sigma^2 = \left\langle [\vec{x}_\eta(t)]^2 \right\rangle -
  [\left\langle \vec{x}_\eta(t) \right\rangle]^2$, by a continuous dark (black online)
  curve for two noise strength values: $\Gamma = 10^{-2}$ [panel {\bf (a)}] and $\Gamma
  = 10^{-1}$ [panel {\bf (b)}], with constant diffusion coefficients (thermal noise) given
  by $h_{ij} = \delta_{ij}\,\Gamma$. The vertical [horizontal] dashed (red online [blue
  online]) curves represents the transient [asymptotic] analytical prediction of
  Eq.~(\ref{eq_Einstein_rel}) [Eq.~(\ref{eq_std_asympt})] for the behaviour of
  $\sigma^2$.}
 \label{fig_BLE_orbits_std}
\end{figure}

For the Brusselator, the asymptotic behaviour of the variance implies that for the
sub-critical values of the parameter ($b < b_c$), any small deviation from the FP in
presence of moderate additive noise keeps, in average, the same asymptotic state. This is
also valid for the limit-cycle situation when parameters are above the critical point
($b > b_c$) if, in the former equations, we use the Floquet exponents and corresponding
time-dependent eigenvectors \cite{Guckenheimer}.

These results [Eqs.~(\ref{eq_Einstein_rel}) and (\ref{eq_std_asympt})] are shown in
Fig.~\ref{fig_BLE_orbits_std} for the particular case of Eq.~(\ref{eq_Einstein_rel_ex})
with $\Gamma = 10^{-2}$ [Fig.~\ref{fig_BLE_orbits_std}{\bf (a)}] and $\Gamma = 10^{-1}$
[Fig.~\ref{fig_BLE_orbits_std}{\bf (b)}]. As it is seen from the results, both analytical
predictions show good agreement with the numerical experiments. Moreover, the asymptotic
\emph{diffusion}, $\sigma^2/\Gamma^2$, as $\Gamma$ is increased (even up to values close
to $\Gamma\sim10^0$) remains constant and identical to the degrees of freedom of the
system, identically to the transient growth rate $D/\Gamma^2 = 2$.

  \subsection{Stochastic Brusselator's Hopf bifurcation}
   \label{sec_Hopf}
In order to analyse how the Hopf bifurcation that the Brusselator exhibits in the
non-stochastic scenario is modified by the presence of additive or multiplicative noise,
we compute the \emph{orbits quadratic difference}, $\Delta^2$,
\begin{equation}
  \Delta^2 = \frac{1}{T}\sum_{t = t^\star}^T \left[ \vec{x}(t) - \left\langle\vec{x}_\eta(t)
   \right\rangle \right]^2\,,
 \label{eq_quad_diff}
\end{equation}
where $\vec{x}(t) = \left( u(t),\,v(t) \right)$ is the deterministic orbit and $\left\langle
\vec{x}_\eta(t)\right\rangle$ is the noise average orbit for the stochastic scenario. This
measure quantifies the distance between the deterministic and the noise average orbit for
each control parameter.

For our numerical experiments, we generate the deterministic orbit and each realisation of
the stochastic orbits from identical initial conditions. Then, a transient of $t^\star =
10^3$ iterations is removed from the orbits to compute the orbit quadratic difference of
Eq.~(\ref{eq_quad_diff}).

\begin{figure}[htbp]
 \begin{center}
  \begin{minipage}{10pc}
   \textbf{(a)}\\
   \includegraphics[width=10pc]{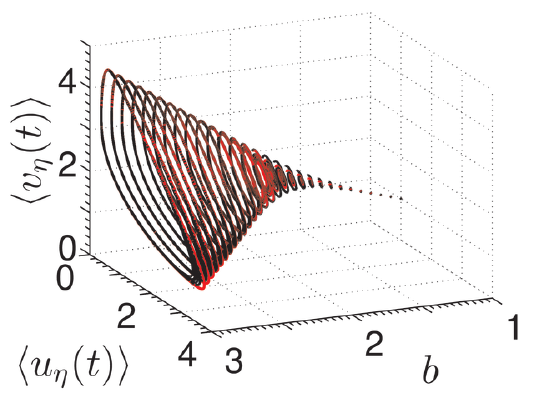}
  \end{minipage}
  \begin{minipage}{10pc}
   \textbf{(b)}\\
   \includegraphics[width=10pc]{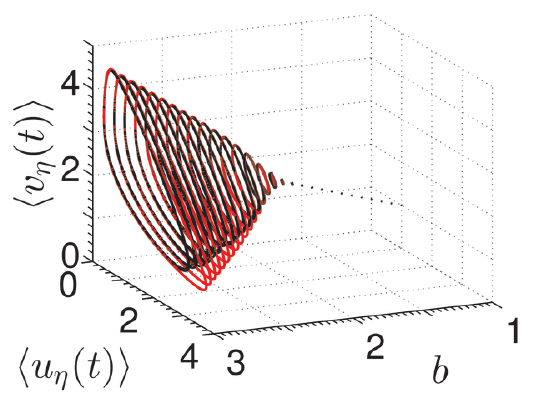}
  \end{minipage}
 \end{center} \vspace{-1pc}
  \caption{(Color online) The left (right) panel shows the Hopf bifurcation that the
  stochastic Brusselator averaged orbits exhibit for thermal (multiplicative) noise with
  constant (linear) diffusion coefficients, $h_{ij}$, as a function of the control parameter
  $b$ for constant $a = 1$. Specifically, the diffusion coefficients are given by $h_{ij} =
  \delta_{ij} \Gamma$ ($h_{ij} = \delta_{ij} \,x_j\,\Gamma$, with $x_1 = u_\eta$ and $x_2 =
  v_\eta$) with $\Gamma = 10^{-2}$. The light (red online) curves correspond to the
  deterministic orbits and the dark (black online) curves correspond to the noise averaged
  orbit in each stochastic scenario [additive noise in panel {\bf (a)} and multiplicative
  noise in panel {\bf (b)}] for $100$ noise realisations. For this $\Gamma$, the
  deterministic and stochastic orbits are very similar, specially for the additive noise
  scenario.}
 \label{fig_SBLE_bifurcation}
\end{figure}

\begin{figure}[htbp]
 \begin{center}
  \begin{minipage}{10pc}
   \textbf{(a)}\\
   \includegraphics[width=10pc]{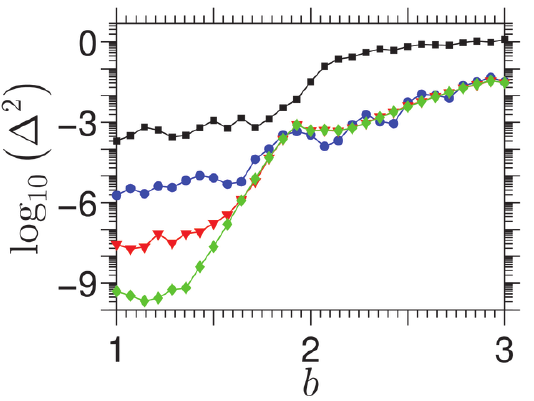}
  \end{minipage}
  \begin{minipage}{10pc}
   \textbf{(b)}\\
   \includegraphics[width=10pc]{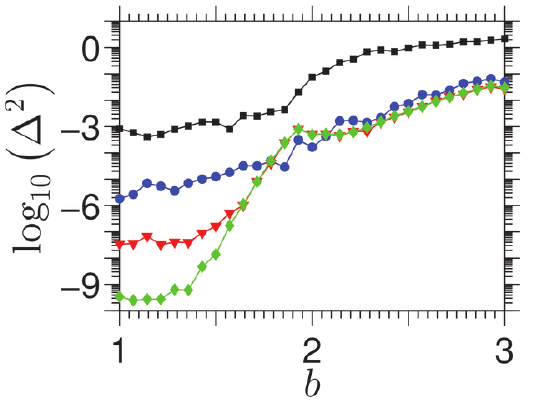}
  \end{minipage}
 \end{center} \vspace{-1pc}
  \caption{(Color online) The orbits quadratic difference, i.e., the time average quadratic
  difference between the deterministic and average stochastic orbit ($\Delta^2 = \frac{1}{T}
  \sum_{t} \left[ \vec{x}(t) - \left\langle\vec{x}_\eta(t) \right\rangle \right]^2$), for
  the additive (left panel) and multiplicative (right panel) noisy Brusselator as the
  control parameter, $b$, is increased for various noise strengths, $\Gamma$. The diffusion
  coefficients are given by $h_{ij} = \delta_{ij}\Gamma$ for panel {\bf (a)} [$h_{ij} =
  \delta_{ij} \,x_j\,\Gamma$, with $x_1 = u_\eta$ and $x_2 = v_\eta$, for panel {\bf (b)}].
  The curves correspond to noise intensities of $\Gamma = 10^{-1}$ (filled --black online--
  squares), $10^{-2}$ (filled --blue online-- circles), $10^{-3}$ (filled --red online--
  triangles), and $10^{-4}$ (filled --green online-- diamonds).}
 \label{fig_SBLE_bif_distan}
\end{figure}

As it is seen from Fig.~\ref{fig_SBLE_bifurcation}, the Hopf bifurcation for the Brusselator
is conserved in the additive [Fig.~\ref{fig_SBLE_bifurcation}{\bf (a)}] and multiplicative
[Fig.~\ref{fig_SBLE_bifurcation}{\bf (b)}] cases for mild noise intensities ($\Gamma =
10^{-2}$). In general, \emph{we observe that the effect of increasing the noise strength
is to reduce the amplitude of the limit-cycle oscillations in the super-critical regime}
($b > b_c$), hence, destroying gradually the Hopf bifurcation. Moreover, the multiplicative
noise realisations generate an even greater decrease in amplitude. Nevertheless, as it is
seen from Fig.~\ref{fig_SBLE_bif_distan}, both stochastic scenarios maintain the bifurcation
type up to noise strengths of $10^{-1}$, where the bifurcation is finally lost.

Besides the determination of the critical noise strength where the Hopf bifurcation is lost,
Fig.~\ref{fig_SBLE_bif_distan} also shows a somehow universal behaviour of the stochastic
system with respect to the deterministic case. In the sub-critical regime, and for parameter
values far from the critical point, the orbits quadratic difference scales linearly with the
noise intensity ($\Delta^2/\Gamma^2 \sim 10^{-1}$). On the other hand, in the supercritical
regime, the orbits quadratic difference collapse under a common curve as a function of the
control parameter. To the best of our knowledge, such behaviour has not been accounted in
previous works.

\section{Discussion}
 \label{sec_conclusions}
In this work we study the Brusselator dynamical behaviour in the absence and presence of
random fluctuations and derive some general expressions for generic stochastic systems.

In the non-stochastic dynamics, all main physical properties of the Brusselator, such as
the spectral values for the equilibrium states, are found and discussed. The inclusion of
thermal and multiplicative noise to the system is first analysed analytically via generic
Stochastic Differential Equations. For the thermal noise scenario, expressions for the
noise average deviation [Eq.~(\ref{eq_mean})], noise average quadratic deviations
[Eq.~(\ref{eq_std})], variance rate growth [Eq.~(\ref{eq_Einstein_rel})], and variance
asymptotic behaviour [Eq.~(\ref{eq_std_asympt})] are derived.

From our numerical experiments, we conclude that the transition that the Brusselator
exhibits from one parameter region, where the chemical concentrations are in a time
independent equilibrium state, to another one, where they oscillate in time, is proved
to be maintained for moderate values of the noise strength ($< 0.1$) in both stochastic
scenarios (additive or multiplicative noise). The character of this transition, which is
Hopf-like for the deterministic evolution, is still observed in the numerical experiments
noise averaged evolutions of the chemical concentrations. Moreover, the analytical
expression for the noise averaged orbit [Eq.~(\ref{eq_mean})] and the variance
[Eq.~(\ref{eq_std})] in the case of additive noise, which we derive in a general framework,
support these findings.

The expression found for the rate at which the variance grows initially
[Eq.~(\ref{eq_Einstein_rel})], is discussed as the Brusselator's diffusion processes and
correlated to the regular Random Walk. Hence, it is accounted as a Einstein diffusion
relationship for the Brusselator. Such analogy is further explored by the derivation of a
general expression for the asymptotic value of the variance of the noisy orbits from the
fixed-point state [Eq.~(\ref{eq_std_asympt})]. Consequently, the stochastic character that
is included into the deterministic Brusselator evolution for the chemical concentrations
accounts for the molecular random fluctuations that the real chemical species involved
in the reaction exhibit.

\section*{Acknowledgements}
The author acknowledges the support of the Scottish University Physics Alliance (SUPA).
The author is in debt with Murilo S. Baptista and Davide Marenduzzo 
for illuminating discussions and helpful comments.


\end{document}